\documentclass[fleqn,twoside]{article}
\usepackage{espcrc2}

\usepackage{graphicx}
\usepackage[figuresright]{rotating}
\usepackage{epsf,epsfig}

\title{Exclusive non-leptonic $B$ meson decays from QCD}

\author{M. Beneke\address[]{Institut f\"ur Theoretische Physik E,
    RWTH Aachen, 
        Sommerfeldstr. 28, D - 52074 Aachen, Germany}%
        \thanks{Talk presented at the 5th KEK Topical Conference 
                ``Frontiers in Flavour Physics'', 
                Tsukuba, Japan, November 2001}}
       
\begin{document}

\begin{abstract}
\noindent
A brief report on the QCD factorization approach to exclusive $B$ decays 
and its applications is presented. In an appendix a subtle issue concerning 
partial integration in the parameterization of infrared-sensitive power 
corrections is clarified.
\vspace{1pc}
\end{abstract}

\maketitle

\vspace*{-6.6cm}
\noindent PITHA 02/06\\
hep-ph/0202056
\vspace*{4.9cm}

\section{INTRODUCTION}
\setcounter{footnote}{0}

Exclusive non-leptonic decays play a crucial part in the on-going 
programme to clarify the sources of CP violation and rare 
flavour-changing phenomena. As regards CP violation this is illustrated 
by the fact that there exists a fair knowledge of the 
length of each of the sides of the unitarity triangle from semi-leptonic 
$B$ decays and $B\bar B$ mixing. Since recently, the angle $\beta$ 
(more precisely, $\sin 2\beta$) is determined fairly accurately 
by the time-dependent CP asymmetry in the non-leptonic decay 
$B_d\to J/\psi K_S$ and related ones. However, the direct measurement 
of the other angle $\gamma$, the phase of $V_{ub}^*$ in the standard 
phase convention, remains a challenge to theory and experiment. 
Although $\gamma$ is determined indirectly through the other 
measurements, the complementary direct determination is important, 
since meson mixing and rare non-leptonic decays could be affected by 
new flavour-changing interactions.

The angle $\gamma$ can be determined from 
decays with some kind of interference of 
$b\to c\bar u D$ (no phase) and $b\to u\bar c D$ (phase $\gamma$) 
tree transitions and their conjugates ($D=d,s$) or from 
decays with interference 
of $b\to u\bar u D$ (tree, phase $\gamma$) and $b\to Dq\bar q$ 
(penguin, phase 0 ($D=s$), $\beta$ ($D=d$)). There exist a few 
methods that allow one to obtain $\gamma$ in a theoretically clean 
way, which means that no calculation of strong interaction effects 
is needed, but these methods are all difficult to realize experimentally 
at present. The more general situation is that some information 
on the decay amplitudes is required to extract $\gamma$ from the 
interference term in branching fractions or CP asymmetries. In 
technical terms one needs to compute the matrix elements 
$\langle f|O_i|\bar B\rangle$, where $O_i$ is an operator 
in the weak effective Hamiltonian and $f$ is a particular 
exclusive final state. 

\section{THEORY OF EXCLUSIVE DECAYS}

\subsection{QCD-improved factorization}

The QCD factorization approach \cite{BBNS1,BBNS2} uses heavy 
quark expansion methods ($m_b\gg\Lambda_{\rm QCD}$) and 
soft-collinear factorization (particle energies $\gg\Lambda_{\rm QCD}$, 
``colour-transparency'') to compute the matrix elements 
$\langle f|O_i|\bar B\rangle$ in an expansion in $1/m_b$ and 
$\alpha_s$. Only the leading term in $1/m_b$ assumes a simple 
form. The basic formula is 
\begin{eqnarray}
\label{fact}
\langle M_1 M_2|O_i|\bar B\rangle 
&\hspace*{-0.2cm}=&\hspace*{-0.2cm} 
F^{B\to M_1}(0)\int_0^1 \!\!du\,T^I(u)
\Phi_{M_2}(u) \nonumber\\[0.0cm]
&&\hspace*{-2.8cm}
+\!\int \!\!d\xi du dv \,T^{II}(\xi,u,v)\,\Phi_B(\xi)\Phi_{M_1}(v) 
\Phi_{M_2}(u),
\end{eqnarray}
valid whenever $M_2$ is a light meson (or, at least in principle, a 
charmonium state) and $M_1$ any light or heavy meson. ($M_1$ is 
the meson that picks up the spectator quark from the $B$ meson. If $M_1$ 
is heavy the second line is higher order in $1/m_b$.) The formula 
shows that there is no long-distance interaction between the
constituents of the meson $M_2$ and the $(B M_1)$ system at leading 
order in $1/m_b$. This is the precise meaning of factorization. In 
particular, the hard scattering kernels $T^{I,II}$ contain 
all rescattering phases at this order and include hard 
interactions with the spectator quark ($T^{II}$). The following 
non-perturbative inputs are needed: $F^{B\to M_1}$ -- a 
$B\to M_1$ form factor; $\Phi_{M_i}$ -- the standard light-cone distribution 
amplitudes for light mesons; $\Phi_B$ -- the light-cone distribution 
amplitude of the $B$ meson \cite{Grozin:1996pq,kpy,Beneke:2000wa,kawa}, 
some aspects of which remain to be understood. The QCD factorization 
approach extends the Brodsky-Lepage approach to exclusive hard processes 
\cite{Lepage:1980fj}, because it shows factorization also when 
soft interactions dominate the $B\to M_1$ form factor, 
as is expected \cite{cz}. 

\subsection{Theory status}

{\em Leading order in $1/m_b$.} Proofs of the factorization formula 
for heavy-light final states exist at two loops \cite{BBNS2} 
and probably to all orders \cite{Bauer:2001cu,Bauer:2001yt}. These 
proofs must be extended to charmless final states by including the 
hard-spectator interaction. In practical applications the calculations 
are currently done at next-to-leading order in perturbation theory. 
There exist theoretical uncertainties from input parameters, which can be 
estimated, the 
most important of which are $|V_{ub}|$ and form factors. The 
theory is presently complete for flavour non-singlet mesons.

{\em Power corrections in $1/m_b$.} Factorization does not and is not 
expected to hold at subleading order. Attempts to compute subleading 
power corrections to hard spectator-scattering in perturbation theory 
usually result in infrared divergences, which signal the breakdown 
of factorization. A subset of power corrections has been parametrized 
or estimated in \cite{BBNS2,khod,BBNS3}. It is common practice in 
applications of perturbative QCD to not include errors from power 
corrections. One would wish to do better here, first because 
non-leptonic decays test fundamental aspects of weak interactions, 
second because some power corrections related to scalar currents 
are enhanced by factors such as 
$m_\pi^2/((m_u+m_d)\Lambda_{\rm QCD})$ \cite{BBNS1}. 
At least these effects should be estimated and included into the 
error budget. This has been attempted in \cite{BBNS3}. All weak 
annihilation contributions belong to this class of effects.

\subsection{Overview of applications}

QCD-improved factorization has been applied to date to the following 
final states:
\begin{itemize}
\item[-] $PP$, where $P$ denotes a light pseudo-scalar non-singlet 
meson \cite{BBNS1,BBNS3,Muta:2000ti,du1,Du:2001hr}. This is the 
application developed furthest.
\item[-] some $PV$ final states, where $V$ denotes a light vector 
meson \cite{Yang:2000xn,Cheng:2000hv}.
\item[-] some $VV$ final states \cite{Cheng:2001aa}. (In the heavy 
quark limit both vector mesons are longitudinally polarized.)
\item[-] $D^{(*)}L$, where $L$ denotes a light pseudo-scalar or 
vector meson \cite{BBNS2,chayD,Neubert:2001sj}.
\item[-] $D^{(*)} X$, where $X$ denotes a scalar or ``exotic'' 
meson \cite{diehl}.
\item[-] $J/\psi K$ \cite{ChayJ,Cheng1}, $J/\psi K^*$ 
\cite{pham,Cheng2}. The factorization approach should apply 
in the limit $m_c\to \infty$, but one expects 
corrections to factorization of order 
$\Lambda_{\rm QCD}/(m_c\alpha_s)\sim 1$ \cite{BBNS2}.
\item[-] Radiative final states $V\gamma$ \cite{BFS,BBg,Ali} and 
$V l^+ l^-$ \cite{BFS} including isospin-breaking effects 
that could cause a different rate for neutral and 
charged $B$ meson decay \cite{KN}.
\item[-] New physics effects due to a modification of the 
$b\to s g$ transition \cite{He} or due to R-parity 
violating supersymmetry \cite{Ghosh}.
\end{itemize}
Of these final states only $D^{(*)}L$ and $PP$ will be discussed below.

\subsection{Other approaches}

{\em Naive factorization} \cite{Wirbel:1985ji}. Naive factorization 
follows from the factorization formula for $D^{(*)}L$, $L_1 L_2$ 
final states, when $\alpha_s\to 0$ and $m_b\to\infty$. QCD-improved 
factorization extends naive factorization in the same sense in which 
the QCD-improved parton model extends the naive parton model.

{\em The ``PQCD'' approach} \cite{Li}. In this approach the decay 
amplitude is calculated in the Brodsky-Lepage formalism. The main 
difference with the QCD-improved factorization approach arises 
through the stronger assumption that Sudakov suppression renders 
the $B\to M_1$ form factor and many power-suppressed effects 
(such as weak annihilation) calculable. The advocates of the 
QCD factorization approach do not agree with the proponents 
of the PQCD approach on conceptual aspects 
of the implementation of Sudakov suppression and on the claim 
that the result is perturbative and accurate \cite{SDG}. 

\section{SELECTED RESULTS}

\subsection{\boldmath $B\to D^{(*)} L$}

\begin{figure}[b]
\vspace*{-0.2cm}
\epsfxsize=7.5cm
\centerline{\epsffile{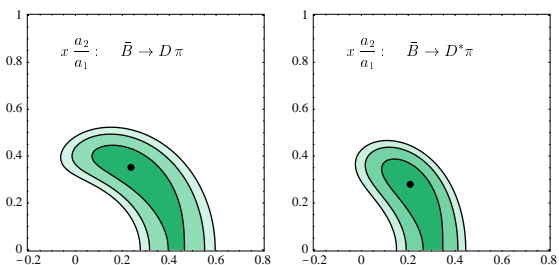}}
\vspace*{-0.6cm}
\caption{\label{fig1}
Determination of $x a_2/a_1$ for $D^{(*)}\pi$ 
final states \cite{Neubert:2001sj}.}
\end{figure}

The phenomenology of these decays is discussed in \cite{BBNS2}. 
QCD-improved factorization explains the experimentally well-established 
fact that the colour-allowed class-I decays to $D^{(*)+}L^-$ are controlled 
by an approximately universal coefficient $|a_1|\simeq 1.05$. On the 
other hand, factorization is not expected to hold for colour-suppressed 
class-II decays to $D^{(*)0}L^0$. The corresponding coefficient 
$a_2$ is therefore depends on the final state. The precise 
nature of $a_2$ depends, however, on the counting scheme. If the $D$ meson is 
considered as heavy ($m_c/m_b$ fixed, $m_b\to\infty$), $x a_2/a_1$ 
is power-suppressed in the heavy quark limit. The power suppression 
is due to 
\begin{equation}
x \equiv \frac{(m_B^2-m_\pi^2)\,f_D\,F_0^{B\to\pi}(m_D^2)}
{(m_B^2-m_D^2)\,f_\pi\,F_0^{B\to D}(m_\pi^2)},
\end{equation}
but with $x \simeq 0.9$ this suppression is only formal. If the $D$ meson 
is considered as light ($m_c$ fixed, $m_b\to\infty$), $x a_2/a_1$ counts 
as order 1 and becomes calculable. (A crude estimate has been given 
in \cite{BBNS2}.) In this case one expects an 
important non-universal contribution from spectator-scattering. In 
both cases $x a_2/a_1$ should be smaller than 1 due to 
colour factors. 

The measurement of $\bar B^0\to D^{(*)0}\pi^0$ allows one for the first time 
to construct the complete isospin triangle for $B\to D^{(*)}\pi$ decays 
and hence to determine the magnitude and rescattering phase of 
$x a_2/a_1$, see Figure~\ref{fig1}. The result is that $|a_2|\simeq 0.4$ 
is sizeable for $\pi D$ final states. Together with $|a_2|\simeq 0.25$ 
for $K J/\psi$ final states this provides clear evidence for the 
non-universality of colour-suppressed decays, but it is 
difficult to interpret this result quantitatively since the theory 
is not expected to be accurate for $\pi^0 D^0$ and $K J/\psi$ 
final states. For further discussion of this result from the 
perspective of QCD factorization see \cite{Neubert:2001sj}.

\subsection{\boldmath $\gamma$ from $B\to \pi\pi,\pi K$}

The possibility to determine the CP-violating angle $\gamma$ 
by comparing the calculation of branching fractions 
into $\pi\pi$ and $\pi K$ final states with the corresponding 
data has been investigated in detail \cite{BBNS3} 
(see also \cite{Du:2001hr}). The branching fractions for the modes 
$B^\mp\to\pi^\mp\pi^0$
  and $B^\mp\to\pi^\mp\bar K^0$, which depend only on a single weak
  phase to very good approximation, are well described by the 
theory. This demonstrates that the
  magnitude of the tree and penguin amplitude is obtained
  correctly, where for the penguin amplitude the 1-loop radiative 
correction is important to reach this conclusion. There is, however, a 
relatively large normalization 
  uncertainty for the $\pi K$ final states, which are sensitive to 
  weak annihilation and the strange quark mass through the scalar
  penguin amplitude. This uncertainty can be partially eliminated by
  taking ratios of branching fractions. The agreement is less 
  good for branching fractions with significant interference of tree
  and penguin amplitudes, if $\gamma$ is assumed to take values 
  around $55^\circ$ as favoured by indirect constraints. This may 
  indicate an interesting alternative determination of $\gamma$, 
but the CP-averaged 
branching fractions sensitive to $\gamma$ also depend on 
the strong phase difference of the two interfering amplitudes. 
This dependence is suppressed as long as the 
phase difference is small since only its cosine enters the calculation. 
With some exceptions (such as the final state $\pi^0\pi^0$) 
QCD-improved factorization predicts that strong phase differences are not 
large, since they are parametrically of order $\alpha_s$ or 
$\Lambda_{\rm QCD}/m_b$, and it therefore also 
  predicts small direct CP asymmetries, up to $(10-15)\%$ (with a 
  preference around $5\%$) for the $\pi^0K^\pm$, $\pi^\pm K^0$ final
  states. The current data seem to favour small 
  CP asymmetries, 
  but they are not yet
  accurate enough to decide upon whether the QCD factorization approach
  allows one to predict strong phases quantitatively.

\begin{figure}[t]
\vspace*{0.0cm}
\epsfxsize=8cm
\centerline{\epsffile{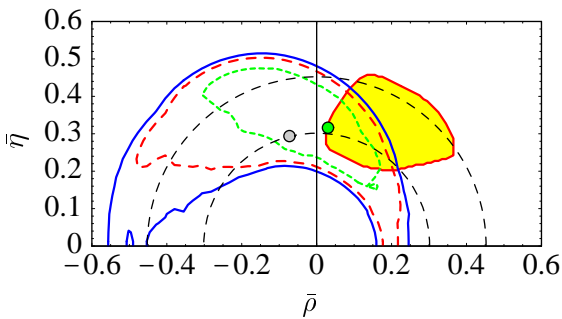}}
\vspace*{-0.6cm}
\caption{\label{fig2}
95\% (solid), 90\% (dashed) and 68\% (short-dashed) confidence level 
contours in the $(\bar\rho,\bar\eta)$ plane obtained from a global 
fit to the CP averaged $B\to\pi K,\pi\pi$ branching fractions, using 
the scanning method as described in 
\cite{Hocker:2001xe}. The right dot shows the
overall best fit, whereas the left dot indicates the best fit for the
default hadronic parameter set. The light-shaded region indicates the region 
preferred by the standard global fit \cite{Hocker:2001xe}, 
excluding the direct measurement 
of $\sin(2\beta)$.}
\end{figure}

The current interpretation of charmless non-leptonic 
$B$ decays can be summarized
by a global fit of the Wolfenstein parameters $\bar \rho$, $\bar \eta$
to the six CP averaged $\pi\pi$, $\pi K$ branching fractions shown in 
Figure~\ref{fig2}. The result is 
consistent with the standard fit 
based on meson mixing and $|V_{ub}|$, 
but shows a preference for larger $\gamma$ or smaller $|V_{ub}|$. 
If the estimate of the theory uncertainty (included in the curves 
in the Figure) is correct, non-leptonic decays together with 
$|V_{ub}|$ from semi-leptonic decays already imply the existence 
of a CP-violating phase of $V_{ub}$ at the 2-3 $\sigma$ level. 

\subsection{\boldmath $\sin(2\alpha)$ or $\gamma$ from 
$B_d(t)\to\pi^+\pi^-$}

\begin{figure}[t]
\epsfxsize=6cm
\centerline{\epsffile{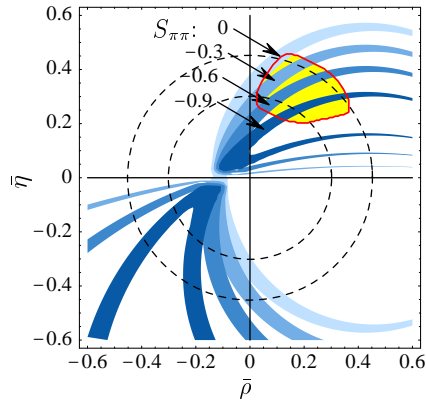}}
\vspace*{-0.8cm}
\caption[dummy]{\label{fig3} Constraint from $S_{\pi\pi}$ in the 
$(\bar\rho,\bar\eta)$ plane.}
\end{figure}

The angle $\alpha$ (or $\gamma$, given the $B\bar B$ mixing phase), 
can be determined from the 
time-dependent CP asymmetry in $B_d\to \pi^+\pi^-$ decay alone, if 
the relative magnitude of the penguin amplitude, $P/T$, can be computed. 
The asymmetry is given by
\begin{eqnarray}
   A_{\rm CP}[\pi\pi](t) 
&\hspace*{-0.2cm}=&\hspace*{-0.2cm} 
  S_{\pi\pi} \sin(\Delta M_{B_d}t) \nonumber\\
&&
   + \,A_{\rm CP}^{\rm dir}[\pi\pi] \cos(\Delta M_{B_d}t),
\end{eqnarray}
where $S_{\pi\pi} = \sin(2\alpha)$, if $P/T=0$, in which case 
the direct CP asymmetry $A_{\rm CP}^{\rm dir}[\pi\pi]$ vanishes. 
The direct CP asymmetry is proportional to the sine of the strong 
phase of $P/T$ and can be used as a phenomenological check of the
computation of $P/T$. Figure~\ref{fig3} displays how a measurement 
of $S_{\pi\pi}$ constrains  
$(\bar\rho,\bar\eta)$, when $P/T$ is 
computed in the QCD factorization approach \cite{BBNS3}. 
The Figure illustrates that even if theoretical (or experimental) 
uncertainties prevent an accurate determination of $\sin(2\alpha)$ 
in this way, the inaccurate result on $\sin (2\alpha)$ still translates into 
a useful constraint in the $(\bar\rho,\bar\eta)$ plane. This reflects 
the fact that other observables do not currently 
constrain $\sin(2\alpha)$ very well.

\section{CONCLUSION}

Interpreting the new data on exclusive two-body decays of $B$ mesons 
in terms of fundamental parameters of the Standard Model requires 
-- at least for the present -- control over strong interaction effects. 
QCD-improved factorization allows us to formulate and discuss 
the problem systematically in the expansion parameters $\alpha_s$ 
and $\Lambda_{\rm QCD}/m_b$. The first comparisons of the theory 
with data on heavy-light and charmless final states are encouraging, 
but the theory may not always be accurate or useful. More data and 
calculations should help understanding where the theory works or fails, 
and if it fails, why it  does. So far no unexpected failures are 
observed. If the estimates of theoretical uncertainties are correct, 
$B\to \pi\pi, \pi K$ together with $|V_{ub}|$ from semi-leptonic 
decay establishes $\gamma\not=0$ at the 2-3$\,\sigma$ level.

\section*{ACKNOWLEDGEMENT}

I wish to thank the organizers for their generous support and 
hospitality and V.M.~Braun and M.~Diehl for comments on the 
manuscript.

\section*{APPENDIX}

In this appendix I address a subtle point in the parameterization 
of infrared sensitive twist-3 corrections to non-leptonic decays. 
The point concerns the legitimacy of integrating by parts in convolution 
integrals with apparently divergent boundary terms. Depending 
on the procedure the result appears to be either a linear or a 
logarithmic infrared divergence. The difference is crucial, since 
a linear divergence compensates the $1/m_b$-suppression and hence 
spoils factorization even in the heavy quark limit, 
while a logarithmic divergence does not.  
The results of the following discussion have been used but have not 
been explained in our previous work \cite{BBNS3}. 

In order to dispose of the need to introduce $B$ meson distribution 
amplitudes I consider an analogous problem that appears in the 
calculation of the pion form factor $\langle \pi^+(p')|
\bar u\gamma_\mu u|\pi^+(p)\rangle$ at large momentum transfer 
$Q^2=-(p'-p)^2$. The light-cone expansion in $1/Q^2$ can be implemented 
by multiplying the on-shell four-quark scattering amplitude 
with a projection operator. For the following discussion it will suffice 
to restrict the projection to the two-particle quark-antiquark 
state and to neglect meson mass corrections.
The coordinate space definitions \cite{Braun:1989iv}
{\small 
\begin{eqnarray*}
\langle \pi^+(p')|\,\bar u_\alpha(y) \, d_\beta(x)|0\rangle
&\hspace*{-0.2cm}=&\hspace*{-0.2cm} 
\frac{i f_\pi}{4} \int_0^1 \!du \,  e^{i (u \, p'\cdot y +
    \bar u p' \cdot x)} 
\nonumber\\[0.1cm]
&& \hspace*{-3.95cm}\times \left\{\!\not \!p' \gamma_5 \, \phi(u)
                 - \mu_\pi \gamma_5 \left( \phi_p(u)
                 - \sigma_{\mu\nu} p'{}^{\mu} 
                 z^\nu \, \frac{\phi_\sigma(u)}{6}\right)
\right\}_{\beta\alpha}
\label{pispacedef}
\end{eqnarray*}
}\hspace*{-0.13cm}
are equivalent to projecting the scattering amplitude with 
{\small 
\begin{eqnarray*}
M_{\beta\alpha} 
&\hspace*{-0.2cm}=&\hspace*{-0.2cm} 
\frac{i f_\pi}{4} 
  \, \bigg\{ \!\not \! p' \gamma_5 \, \phi(u)  - 
   \mu_\pi \gamma_5 \bigg(  \phi_p(u) 
  \nonumber\\[0.2em]
&& \hspace*{-1.3cm}
       +\,i \sigma_{\mu\nu}  p'{}^\mu \, \frac{\phi_\sigma(u)}{6}
       \, \frac{\partial}{\partial
         k_{\nu}} \bigg)
 \bigg\}_{\beta\alpha}
\label{pimeson2}
\end{eqnarray*}
}\hspace*{-0.13cm}
and to integrate over $u$ from 0 to 1. (All quark-antiquark operators
are assumed to be colour singlets.) I will refer to this 
projection as ``P1''. Here $k$ is the momentum of 
the quark in the pion and the antiquark momentum is $p'-k$. The derivative 
can be decomposed into a transverse derivative, a term proportional to 
$p'_\nu$, which does not contribute, and a term 
proportional to $p_\nu$. Using $u=p\cdot k/p'\cdot p$ one performs an 
integration by parts on this term and neglecting boundary terms (because 
$\phi_\sigma(u)$ vanishes at 0 and 1), one obtains the projection 
``P2'':
{\small 
\begin{eqnarray*}
M_{\beta\alpha} 
&\hspace*{-0.2cm}=&\hspace*{-0.2cm} 
\frac{i f_\pi}{4} 
  \, \bigg\{ \!\not \! p' \gamma_5 \, \phi(u)  - 
   \mu_\pi \gamma_5 \bigg(  \phi_p(u) 
  \nonumber\\[0.2em]
&& \hspace*{-1.3cm}
       - \,i \sigma_{\mu\nu}\,  \frac{p'{}^\mu  p^\nu}{p'\cdot p} \,
       \frac{\phi^\prime_\sigma(u)}{6}
 +i \sigma_{\mu\nu}  p'{}^\mu \, \frac{\phi_\sigma(u)}{6}
       \, \frac{\partial}{\partial
         k_\perp{}_{\nu}} \bigg)
 \bigg\}_{\beta\alpha}
\end{eqnarray*}
}\hspace*{-0.1cm}
The calculation of the form factor then gives
{\small 
\begin{eqnarray*}
\langle \pi^+(p')|
\bar u\gamma_\mu u|\pi^+(p)\rangle
&\hspace*{-0.2cm}=&\hspace*{-0.2cm} 
\frac{4 g_s^2 f_\pi^2}{9 Q^2} \int_0^1 dudv   \nonumber\\[0.2em]
&& \hspace*{-2.5cm}
\times\bigg\{\frac{
\phi(u)\phi(v)}{2\bar u\bar v}\,(p+p')_\mu + 
\frac{\mu_\pi^2}{Q^2}\,T_4\bigg\},
\label{ff1}
\end{eqnarray*}
}\hspace*{-0.13cm}
where the first term in the bracket represents the leading-twist 
result and $T_4$ a twist-4 correction, which is the quantity of 
interest here.

The explicit form of $T_4$ depends on whether P1 or P2 is used for 
the calculation. With P1 the result is
{\small 
\begin{eqnarray*}
\phi_p(u)\bigg[\bigg\{-\frac{1}{\bar u\bar v}\,\phi_p(v)+
\frac{1}{\bar u\bar v^2}\,\frac{\phi_\sigma(v)}{6}\bigg\}\,p_\mu&&
\nonumber\\[0.2em]
&&\hspace*{-5cm} 
+\,\bigg\{-\frac{1}{\bar u\bar v^2}\,\phi_p(v) - 
\frac{2}{\bar u\bar v^3}\,\frac{\phi_\sigma(v)}{6}\bigg\}\,p'_\mu
\bigg] \nonumber\\[0.2em]
&&\hspace*{-5cm} 
+\, \left(u\leftrightarrow v,p\leftrightarrow p' \right),
\end{eqnarray*}
}\hspace*{-0.1cm}
while using P2 leads to 
{\small 
\begin{eqnarray*}
\phi_p(u)\bigg[\bigg\{-\frac{1}{\bar u\bar v}\,\phi_p(v)
+\frac{1}{\bar u\bar v}\,\frac{\phi'_\sigma(v)}{6}+
\frac{2}{\bar u\bar v^2}\,\frac{\phi_\sigma(v)}{6}\bigg\}\,p_\mu&&
\nonumber\\[0.2em]
&&\hspace*{-6.7cm} 
+\,\bigg\{-\frac{1}{\bar u\bar v^2}\,\phi_p(v) +
\frac{1}{\bar u\bar v^2}\,\frac{\phi'_\sigma(v)}{6}\bigg\}\,p'_\mu
\bigg] \nonumber\\[0.2em]
&&\hspace*{-6.7cm} 
+\, \left(u\leftrightarrow v,p\leftrightarrow p' \right).
\end{eqnarray*}
}\hspace*{-0.17cm}
The two results are identical, when an integration by parts is 
performed and when the boundary terms are zero. This does 
not seem to be justified, however, since the boundary terms 
are non-zero, in fact infinite, when the asymptotic distribution 
amplitudes $\phi_p(w)=1$, $\phi_\sigma(w) = 6 w\bar w$ are 
inserted.\footnote{In the present approximation that neglects 
quark-anti\-quark-gluon amplitudes the asymptotic distribution 
amplitudes are exact and follow from the equation of motion constraints. 
A possible caveat here is that the solutions given above 
also involve an integration by parts in which boundary terms 
should be zero even when the equation of motion  
is inserted into any Green function. The following discussion 
shows that this assumption can be made.} Not performing the 
integration by parts, it is seen that the result from P1 is 
linearly divergent as $v\to 1$, while the result from P2 exhibits 
only a logarithmic divergence. Since non-perturbative infrared 
effects regulate the divergence at $\bar v\approx \Lambda/Q$ 
(when the quark momentum becomes of order $\Lambda$), the 
P1-result is of order $Q/\Lambda$, parametrically larger  
than the P2-result which is only of order $\ln Q/\Lambda$. This 
can be made more explicit by regulating the convolution integrals 
through introducing a gluon mass $\lambda$. The main effect 
of this is to replace $\bar u\bar v$ by $\bar u\bar v+\xi$ where 
$\xi=\lambda^2/Q^2$. The 
convolution integrals can then be done with the two 
results:
{\small 
\begin{eqnarray*}
\int_0^1 dudv\, T_4(P1)&\hspace*{-0.2cm}=&\hspace*{-0.2cm} 
\left(-\frac{1}{\xi}-2 \,\mbox{Li}_2\!\left(-\frac{1}{\xi}\right)\right)\,
(p+p')_\mu \nonumber\\[0.2em]
\int_0^1 dudv\, T_4(P2)&\hspace*{-0.2cm}=&\hspace*{-0.2cm} 
-2 \,\mbox{Li}_2\!\left(-\frac{1}{\xi}\right)\,(p+p')_\mu \nonumber
\end{eqnarray*}
}\hspace*{-0.1cm}
The difference $-1/\xi$ can be attributed to a linearly divergent 
boundary term
{\small 
\begin{displaymath}
\lim_{\bar v\to 0} \frac{1}{\bar v(\bar u\bar v+\xi)}\,
\frac{\phi_\sigma(v)}{6} = \frac{1}{\xi}.
\end{displaymath}
}\hspace*{-0.13cm}

Now it is clear that was has gone wrong here is that one divides 
0 by 0. The numerator is $\phi_\sigma(1)=0$, which is always correct, 
but the denominator zero arises from the virtuality of a quark 
propagator, which approaches zero and which is not regularized by 
introducing a gluon mass. To make the calculation well-defined all 
small-virtuality regions must be properly regularized. One possibility 
is to assign a small mass to the quarks as well. Denoting the corresponding 
parameter by $\eta$, this 
leads to the boundary term 
{\small 
\begin{displaymath}
\lim_{\bar v\to 0} \frac{1}{(\bar v+\eta)(\bar u\bar v+\xi)}\,
\frac{\phi_\sigma(v)}{6} = 0.
\end{displaymath}
}\hspace*{-0.4cm}
It is not difficult to see that in this particular case the 
amplitude has a finite limit for $\eta\to 0$ (but $\xi$ 
still non-zero) after the convolution integrals are performed. 
The only effect of the complete regularization is therefore 
to make all integrations by parts well-defined and to eliminate 
the spurious linearly divergent boundary term. We can therefore 
conclude that the correct result is 
{\small 
\begin{eqnarray*}
T_4(P2)&\hspace*{-0.2cm}=&\hspace*{-0.2cm} \frac{2}{\bar u\bar v}\,
(p+p')_\mu, \nonumber
\end{eqnarray*}
}\hspace*{-0.1cm}
which exhibits a 
double logarithmic divergence, one from each convolution integral. 
This coincides with the result for the chirally-enhanced twist-4 
corrections obtained long ago by Geshkenbein and Terentev 
\cite{Geshkenbein:qn}.

The discussion can be analogously repeated for non-leptonic 
$B$ meson decays, where the problem appears in the computation of 
$1/m_b$-suppressed terms. The conclusion is again that integration 
by parts is allowed and that the correct result is obtained directly 
with the projector P2. On the other hand using P1 requires one 
to regularize carefully all propagators that can go close to 
mass-shell near the endpoints of the integration region. 

There is a deeper reason why the infrared divergence in the twist-4 
correction to the pion form factor should not be 
power-like. There exist two basic scattering mechanisms that contribute 
to the form factor. One is the hard scattering mechanism, in which 
all components of a given Fock state of the pion participate in a 
hard interaction. The other is a soft-overlap contribution, in which 
case a number of Fock components with small longitudinal momentum 
fraction does not participate in a hard interaction. A complete 
asymptotic expansion of the form factor should combine both mechanisms 
consistently. Recent work on the analogous problem for the 
$B\to\pi$ form factor \cite{Beneke:2000wa} together with the structure
of the radiatively corrected QCD sum rule for the $B\to \pi$ form
factor \cite{Bagan:1997bp} suggests the following factorization formula 
that incorporates the hard scattering and soft-overlap 
contribution:
{\small 
\begin{eqnarray*}
F(q^2) 
&\hspace*{-0.2cm}=&\hspace*{-0.2cm} 
\frac{1}{Q^2}\, T_2*\phi_2*\phi_2 \\
&& \hspace*{-1cm}
+\,C\,\xi(q^2,\mu)+\frac{1}{Q^4} \,T_4^{\rm reg}(\mu)*\phi_3*\phi_3
+O(1/Q^6)
\end{eqnarray*}
}\hspace*{-0.2cm}
Here $\phi_2$ and $\phi_3$ represent twist-2 and twist-3 light-cone 
distribution amplitudes and $\xi(Q^2,\mu)$ a soft-overlap form factor which 
scales as $1/Q^4$ and which can be defined in an effective theory 
in which all hard scatterings are integrated out. $T_4^{\rm reg}(\mu)$ 
denotes the hard-scattering kernel at the twist-4 level discussed 
above except that the endpoint divergence is factorized into the 
soft-overlap form factor, such that both terms depend on a 
factorization parameter $\mu$. If this (conjectured) factorization 
formula is to be correct, the 
endpoint contribution of the twist-4 term can be at most logarithmically 
enhanced.


\begin{thebibliography}{99}

\bibitem{BBNS1}
M.~Beneke, G.~Buchalla, M.~Neubert and C.~T.~Sachrajda,
Phys.\ Rev.\ Lett.\  {\bf 83}, 1914 (1999).

\bibitem{BBNS2}
M.~Beneke, G.~Buchalla, M.~Neubert and C.~T.~Sachrajda,
Nucl.\ Phys.\ B {\bf 591}, 313 (2000).

\bibitem{Grozin:1996pq}
A.~G.~Grozin and M.~Neubert,
Phys.\ Rev.\ D {\bf 55}, 272 (1997).

\bibitem{kpy}
G.~P.~Korchemsky, D.~Pirjol and T.~M.~Yan,
Phys.\ Rev.\ D {\bf 61} (2000) 114510.

\bibitem{Beneke:2000wa}
M.~Beneke and T.~Feldmann,
Nucl.\ Phys.\ B {\bf 592} (2001) 3.

\bibitem{kawa}
H.~Kawamura, J.~Kodaira, C.~F.~Qiao and K.~Tanaka,
Phys.\ Lett.\ B {\bf 523} (2001) 111.

\bibitem{Lepage:1980fj}
G.~P.~Lepage and S.~J.~Brodsky,
Phys.\ Rev.\ D {\bf 22} (1980) 2157.

\bibitem{cz}
V.~L.~Chernyak and I.~R.~Zhitnitsky,
Nucl.\ Phys.\ B {\bf 345} (1990) 137.

\bibitem{Bauer:2001cu}
C.~W.~Bauer, D.~Pirjol and I.~W.~Stewart,
Phys.\ Rev.\ Lett.\  {\bf 87} (2001) 201806.

\bibitem{Bauer:2001yt}
C.~W.~Bauer, D.~Pirjol and I.~W.~Stewart,
[hep-ph/0109045].

\bibitem{khod}
A.~Khodjamirian,
Nucl.\ Phys.\ B {\bf 605} (2001) 558.

\bibitem{BBNS3}
M.~Beneke, G.~Buchalla, M.~Neubert and C.~T.~Sachrajda,
Nucl.\ Phys.\ B {\bf 606} (2001) 245.

\bibitem{Muta:2000ti}
T.~Muta, A.~Sugamoto, M.~Z.~Yang and Y.~D.~Yang,
Phys.\ Rev.\ D {\bf 62} (2000) 094020.

\bibitem{du1}
D.~s.~Du, D.~s.~Yang and G.~h.~Zhu,
Phys.\ Lett.\ B {\bf 488} (2000) 46.

\bibitem{Du:2001hr}
D.~s.~Du, H.~u.~Gong, J.~f.~Sun, D.~s.~Yang and G.~h.~Zhu,
[hep-ph/0108141].

\bibitem{Yang:2000xn}
M.~Z.~Yang and Y.~D.~Yang,
Phys.\ Rev.\ D {\bf 62} (2000) 114019.

\bibitem{Cheng:2000hv}
H.~Y.~Cheng and K.~C.~Yang,
Phys.\ Rev.\ D {\bf 64} (2001) 074004.

\bibitem{Cheng:2001aa}
H.~Y.~Cheng and K.~C.~Yang,
Phys.\ Lett.\ B {\bf 511} (2001) 40.

\bibitem{chayD}
J.~Chay,
Phys.\ Lett.\ B {\bf 476} (2000) 339.

\bibitem{Neubert:2001sj}
M.~Neubert and A.~A.~Petrov,
Phys.\ Lett.\ B {\bf 519} (2001) 50.

\bibitem{diehl}
M.~Diehl and G.~Hiller,
JHEP {\bf 0106} (2001) 067.

\bibitem{ChayJ}
J.~Chay and C.~Kim,
[hep-ph/0009244].

\bibitem{Cheng1}
H.~Y.~Cheng and K.~C.~Yang,
Phys.\ Rev.\ D {\bf 63} (2001) 074011.


\bibitem{pham}
X.~S.~Nguyen and X.~Y.~Pham,
[hep-ph/0110284].


\bibitem{Cheng2}
H.~Y.~Cheng, Y.~Y.~Keum and K.~C.~Yang,
[hep-ph/0111094].


\bibitem{BFS}
M.~Beneke, T.~Feldmann and D.~Seidel,
Nucl.\ Phys.\ B {\bf 612} (2001) 25.

\bibitem{BBg}
S.~W.~Bosch and G.~Buchalla,
Nucl.\ Phys.\ B {\bf 621} (2002) 459.


\bibitem{Ali}
A.~Ali and A.~Y.~Parkhomenko,
[hep-ph/0105302].

\bibitem{KN}
A.~L.~Kagan and M.~Neubert,
[hep-ph/0110078].


\bibitem{He}
X.~G.~He, J.~Y.~Leou and J.~Q.~Shi,
Phys.\ Rev.\ D {\bf 64} (2001) 094018.


\bibitem{Ghosh}
D.~K.~Ghosh, X.~G.~He, B.~H.~McKellar and J.~Q.~Shi,
[hep-ph/0111106].


\bibitem{Wirbel:1985ji}
M.~Wirbel, B.~Stech and M.~Bauer,
Z.\ Phys.\ C {\bf 29} (1985) 637.

\bibitem{Li}
See the review by H.-n.~Li in these proceedings 
[hep-ph/0110365].

\bibitem{SDG}
S.~Descotes-Genon and C.~T.~Sachrajda,
[hep-ph/0109260].

\bibitem{Hocker:2001xe}
A.~H\"ocker, H.~Lacker, S.~Laplace and F.~Le Diberder,
Eur.\ Phys.\ J.\ C {\bf 21}, 225 (2001).

\bibitem{Braun:1989iv}
V.~M.~Braun and I.~E.~Filyanov,
Z.\ Phys.\ C {\bf 48} (1990) 239.

\bibitem{Geshkenbein:qn}
B.~V.~Geshkenbein and M.~V.~Terentev,
Phys.\ Lett.\ B {\bf 117} (1982) 243.

\bibitem{Bagan:1997bp}
E.~Bagan, P.~Ball and V.~M.~Braun,
Phys.\ Lett.\ B {\bf 417} (1998) 154.

\end{thebibliography}
\end{document}